# Probing the graphene/substrate interaction by electron tunneling decay

*V. Carnevali [a,1], A. Sala [a,b,\*], P. Biasin [a], M. Panighel [b], G. Comelli [a,b], M. Peressi [a,\*], C. Africh [b]*

[a] *Department of Physics, University of Trieste, via Valerio 2, 34127, Trieste, Italy*
[b] *CNR-IOM, Laboratorio TASC, S.S. 14 km 163.5, Basovizza, 34149, Trieste, Italy*

\* Corresponding authors: A. Sala +39 040 375 6452 sala@iom.cnr.it (experiment), M. Peressi +39 040 2240 242 peressi@units.it (theory)

[1] Present address: Institute of Chemical Science and Engineering, EPFL, 1015 Lausanne, Switzerland

ABSTRACT: *The electronic properties of graphene can be modified by the local interaction with a selected metal substrate. To probe this effect, Scanning Tunneling Microscopy is widely employed, particularly by means of local measurement via lock-in amplifier of the differential conductance and of the field emission resonance. In this article we propose an alternative, reliable method of probing the graphene/substrate interaction that is readily available to any STM apparatus. By testing the tunneling current as function of the tip/sample distance on nanostructured graphene on Ni(100) and Ir(100), we demonstrate that I(z) spectroscopy can be quantitatively compared with Density Functional Theory calculations and can be used to assess the nature of the interaction between graphene and substrate. This method can expand the capabilities of standard STM systems to study graphene/substrate complexes, complementing standard topographic probing with spectroscopic information.*

1. INTRODUCTION

Since the discovery of graphene, Scanning Tunneling Microscopy (STM) has been widely employed for the characterization of two-dimensional materials. In addition to its capability to deliver topographic images with atomic resolution, when used in spectroscopic mode it offers direct access to the local electronic structure, e.g. by probing the differential tunneling conductance $dI/dV$ at constant tip-sample separation $z$[1,2] or the field emission resonance at constant tunneling current[3,4]. It is thus possible to obtain very valuable information also on the morphology of defects[5,6] and on the magnitude of the graphene/substrate interaction[7], which play a fundamental role in determining the electronic and magnetic properties. However, the spectroscopic techniques mentioned above require a dedicated apparatus, most commonly a lock-in amplifier, and often liquid He cooling to avoid thermal noise. In this article we propose a different and simpler approach, where the graphene/substrate interaction can be quantitatively determined by employing a method that is readily available to any basic STM system, without the need of any additional electronic device. More specifically, we show that the analysis of the tunneling current decay ($I$) as a function of $z$ can accurately indicate whether graphene deposited on a surface preserves or not its linearly dispersed electronic π-bands. Our experimental findings, corroborated by theoretical calculations, define a procedure that can be applied in principle to a great variety of graphene/substrate systems.

## 2. MATERIALS AND METHODS

STM experiments were performed using an Omicron Low-Temperature Scanning Tunneling Microscope (STM) kept at 77 K and at a base pressure below $7 \times 10^{-11}$ mbar. I(z) spectroscopy was extracted with direct measurement of the tunneling current while the sample-tip distance was linearly increased by 500 pm. The starting point was set by standard current and bias parameters (see text below). The Ni(100) surface was initially cleaned by iterated cycles of Ar+ sputtering at 1.5 keV and annealing at 870 K. Graphene was grown on Ni(100) kept at 830 K via Chemical Vapor Deposition (CVD) of ethylene, using a well-established recipe[8–10], i.e. 20' exposure at $5 \times 10^{-6}$ mbar and 180' exposure at $5 \times 10^{-7}$ mbar. Slow cooling in the 680-580 K range right after the CVD process favored accumulation of nickel carbide at the interface. Ir(100) was cleaned with iterated cycles of Ar+ sputtering at 1.5 keV, and annealing in oxygen (partial pressure $1 \times 10^{-6}$ mbar) from 570 K to 1200 K. After a final flash in UHV at 1200 K, the sample was kept at 1030 K to grow graphene using a well-established recipe[11], i.e. via CVD of ethylene (10 L at partial pressure $5 \times 10^{-8}$ mbar).

## 3. THEORY/CALCULATION

Density Functional Theory (DFT) simulations were performed with the Quantum ESPRESSO suite of codes[12], using plane-wave basis set and Generalized Gradient Approximation for the

exchange– correlation functional in the Perdew–Burke–Ernzerhof parametrization[13], and van der Waals interactions with the semiempirical DFT-D approach[14]. The graphene/Ni(100) system was modeled with a periodically repeated slab geometry in a simulation cell that allows to accommodate the (2x2) clock reconstruction in the carbide covered region and an optimized corrugated graphene structure in the lateral direction, including chemisorbed, physisorbed and lifted regions. A more detailed description of the simulation cell and other technical details can be found in Ref. [8]. The graphene/Ir(100) system was modeled in a periodically repeated supercell similar to that described in Ref. [11]. For both systems, the vacuum spacing between the two slab replicas was set at least to 20 Å in order to correctly catch the exponential decay of the local density of states. This made the calculations very demanding, in particular for the graphene/Ir(100) system whose supercell contains 325 atoms. The postprocessing included the calculations of charge density differences, local density of states, work function. The work function was computed as the difference between the vacuum level and the Fermi energy of the system.

## 4. RESULTS

### 4.1 Sample morphology

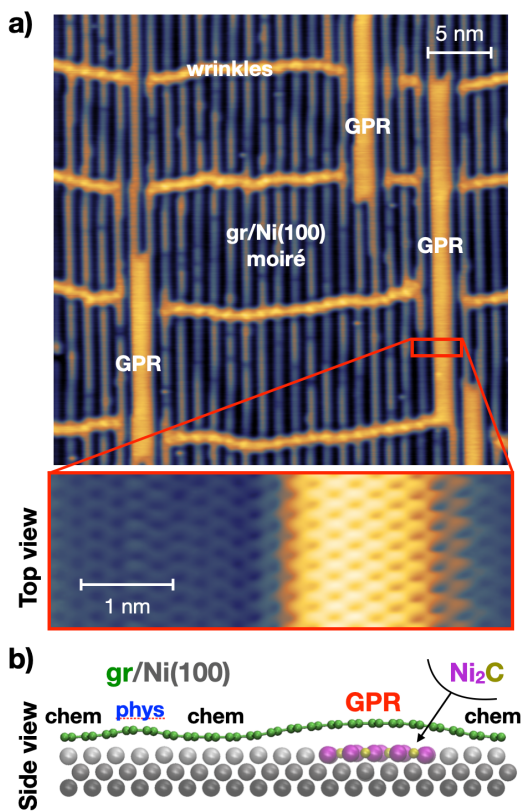

***Figure 1 (1-column, color): atomic-scale topography of graphene/Ni(100).*** *a) STM topographic image (and atomically-resolved zoom) of the as-prepared graphene/Ni(100) surface at 77 K. As previously reported*[8], *graphene presents a 1D moiré modulation interrupted by zigzag-edged pseudo-ribbons (GPR) and wrinkles. $V_B$ = -0.3 V, I = 0.5 nA. b) Side view of the model structure of the zoomed image. In the lateral direction the cell contains 19 Ni unit surface cells and 11 armchair graphene periods*[8]. *The 1D moiré produces alternate lanes of chemisorbed (chem, black label) and physisorbed (phys, blue label) carbon atoms. Lifted pseudo-ribbons (GPR, red label) are produced by interfacial nickel carbide underneath. The color code of the atoms is indicated in the taglines.*

Single-layer graphene grown on Ni(100), is a prototypical system where areas characterized by markedly different interaction with the substrate coexist at a nanometer scale[8]. Typically, on Ni surfaces graphene π-orbitals strongly interact with the d-orbitals of the interfacial nickel atoms, resulting in a narrow separation between the graphene and the metal[9] and, for the former, in the opening of a band gap with a ~ 2.7 eV shift of the π-band[8,15]. In addition, on the Ni(100) surface graphene can be also modulated by the one-dimensional (1D) moiré pattern created by the coincidence of one graphene lattice vector with that of the clean substrate. This results in a wavy, corrugated structure composed by alternated lanes of moiré trenches and ridges - which we label respectively as chemisorbed and physisorbed, in compliance with Ref. [9] - where the graphene/substrate separation ranges from 1.9 Å to 2.9 Å. Moreover, by thermally controlling the amount of carbon atoms segregated at the graphene/substrate interface, it is possible to further locally weaken the interfacial interaction, creating nanometer-sized patches of lifted graphene characterized by restored electronic properties and a graphene/substrate separation of ~ 3.3 Å. STM topography of such weakly interacting stripes embedded in an interacting graphene sheet (**Figure 1a**) shows that the 1D moiré pattern of graphene/Ni(100) is interrupted by lifted one-dimensional stripes (labeled as graphene 'pseudo-ribbons', GPRs) aligned with the moiré, and by occasional wrinkles perpendicular to GPRs[8]. As modeled in **Figure 1b**, the key for this local detachment of graphene is the formation of $Ni_2C(100)$ patches at the graphene/substrate interface, accumulating under a chemisorbed lane of graphene and locally breaking the interaction between the Ni d-orbitals and the graphene π orbitals. In our previous publication we demonstrated that GPRs possess a band structure comparable with weakly interacting graphene nanoribbons of similar size, i.e. the π-band is restored and distorted only by lateral quantum confinement. In synthesis, this system presents in the same graphene sheet three adjacent areas characterized by significantly different electronic structures and morphologies, resulting from different strengths in the graphene/substrate interaction: chemisorbed, physisorbed and GPRs.

We therefore chose this system as a benchmark to demonstrate the possibility of using *I(z)* spectroscopy to unveil the characteristics of the graphene/substrate interaction even at the nanometer scale.

4.2 *I(z) decay*

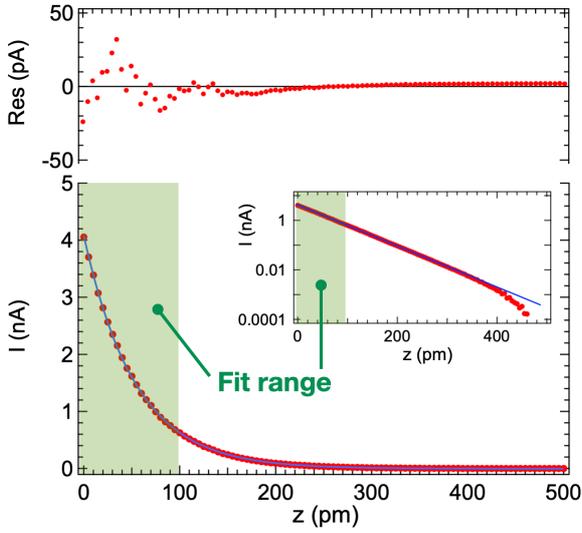

***Figure 2 (1-column, color): I(z) decay.*** *Plot of the tunneling current decay along the z direction (surface normal), together with an exponential fit (blue line) and the fit residuals (on top). The inlay contains the same graph in logarithmic scale. The green regions indicate the proposed fit range. The integration time was 10 ms per point, yielding a 101-points spectrum in ~ 1 s.*

We measured the tunneling current *I* as a function of the tip/sample distance *z* at fixed lateral position. The well-known one-dimensional model for the tunneling barrier introduced by Simmons[16] expresses the tunneling current *I* between two electrodes kept at a bias *V* and placed at a distance *z* as

$$I(z,V) \propto \int_0^{eV} \rho_S(E)\rho_T(E-eV)\,\mathrm{T}(z,E,V)dE \qquad \text{Eq. (1)}$$

Here, $\rho_S$ and $\rho_T$ denote respectively the sample and tip density of states (DOS) and *T* is the Wenzen-Kramer-Brillouin (WKB) tunneling transmission function[17–22]

$$\text{T}(z, E, V) \propto exp\left(-2z\frac{\sqrt{2m}}{\hbar}\sqrt{\emptyset + \frac{eV}{2} - E}\right) = exp\left(-2kz\right) \qquad \text{Eq. (2)}$$

with $\emptyset$ being the apparent tunneling barrier height and *m* the free electron mass. Therefore, $I(z)$ is proportional to $exp(-2kz)$, where $k$ depends on the bias $V$. **Figure 2** shows a typical experimental decay of the tunneling current as a function of $z$, together with the corresponding numerical fit with a negative exponential curve, calculated over a range of 500 pm. As can be seen by the fit residual plot on top and the semi-log graph of the same decay in the inlay, the exponential fit cannot describe $I(z)$ in the whole $z$ range: for intermediate distances the residuals show an overestimation of the current, while at the farthest the fainting signal loses its exponential character as it goes beyond the detection limit of the experimental system (few pA). To strengthen the fit reliability, we have therefore chosen to limit the fit range to the first 100 pm (green transparent area), where the $I(z)$ curve in logarithmic scale has a linear behavior. $I(z)$ spectra were collected by firstly positioning the tip over the points of interest and then moving the tip backward and forward in the $z$ direction, while recording the tunneling current. The exponential decay constant $k$ was finally extracted from the backward/forward averaged spectra.

An estimate of the decay constant can also be obtained by DFT simulations. According to the simplified approach by Tersoff and Hamann[23], for small biases $V$ the STM tunneling current I depends only on the local DOS $\rho_S(r, E)$ of the sample at the position $r$ of the tip

$$I(r, V) \propto \int_0^{eV} \rho_S(r, E_F + E)\, dE \qquad \text{Eq. (3)}$$

where $E_F$ is the Fermi energy of the sample and, at variance with Eq. (1), $\rho_S$ and *I* explicitly depend on the 3D position $r=(x,y,z)$ (and not just on the distance $z$) to account for possible in-plane changes.

The local DOS of the sample $\rho_S$ is determined by the Kohn-Sham orbitals $\psi_\alpha(r)$:

$$\rho_S(r, E) = \sum_\alpha \delta(E - E_\alpha)\, |\Psi_\alpha(r)|^2. \qquad \text{Eq. (4)}$$

Since $|\Psi_\alpha(r)|^2 \propto e^{-2k(V)z}$ in the direction $z$ perpendicular to the surface of the sample, the predicted values for the tunneling current decay constants k(V) were obtained from the exponential decay of the local DOS at fixed *(x,y)* and varying *z* in the 2.4-3.0 Å range, measured from the corrugated graphene layer.

The theoretical simulations of the investigated system, i.e. graphene 'pseudoribbons' on Ni(100), were carried out by using the atomic model presented in Ref. [8], where a complete description of this complex structure is presented. Therefore, in the following we will concentrate only on the results concerning the tunneling current decay. **Figure 3** shows the experimental values of *k* obtained at various locations along the profile shown in Figure 1b, which crosses different relevant areas (chemisorbed, physisorbed and GPR). Measurements are reported for three different bias values, -0.3 V, +0.9 V and +2.0 V and compared to DFT calculations. At all biases, the experimental *I(z)* spectroscopy detects a substantial increase of *k* on GPR and a slight increase on physisorbed lane with respect to the chemisorbed areas. In all three cases this behavior is well reproduced by theory, after a mild Gaussian smoothing that reduces the computational noise. At lower biases (-0.3 V and +0.9 V), the agreement is also quantitative, while for larger bias (+2.0 V), the simulation foresees a larger modulation with respect to the experiment. This discrepancy is likely due to the different tip starting position: while in the simulation the *z* range is kept constant, in the experimental conditions a large bias value yields a very high tunneling current, so that to avoid saturation the initial tip position must be retracted. Thus, the experimental setup forcibly probes only the tail of the tunnel decay, where the effective decay rate is systematically underestimated (see Figure S1 in the Supporting Material for further details). The graph in **Figure 4** displays the averaged *k* values obtained experimentally and theoretically over positions characteristic of the three main graphene regions – chemisorbed (black), physisorbed (blue) and GPR (red). The theoretical decay constant calculated for a freestanding, flat single-layer graphene sheet (1.79±0.07 Å$^{-1}$, green dashed line and box) is also shown for comparison. Such value represents the ideal case, i.e. in absence of any interaction with a substrate. Even a very weak interaction with a substrate can induce shifts in the graphene Fermi level with respect to the Dirac point[24], thus systematically altering the tunneling current decay to lower values. There is a remarkable qualitative agreement between experiment and theory: the less the graphene interacts with the substrate, the more the *k* value increases and leans toward the theoretical freestanding value. Moreover, for smaller bias values, when the tip senses the surface at distance comparable to the one used for the theoretical calculations, there is also a fair quantitative agreement.

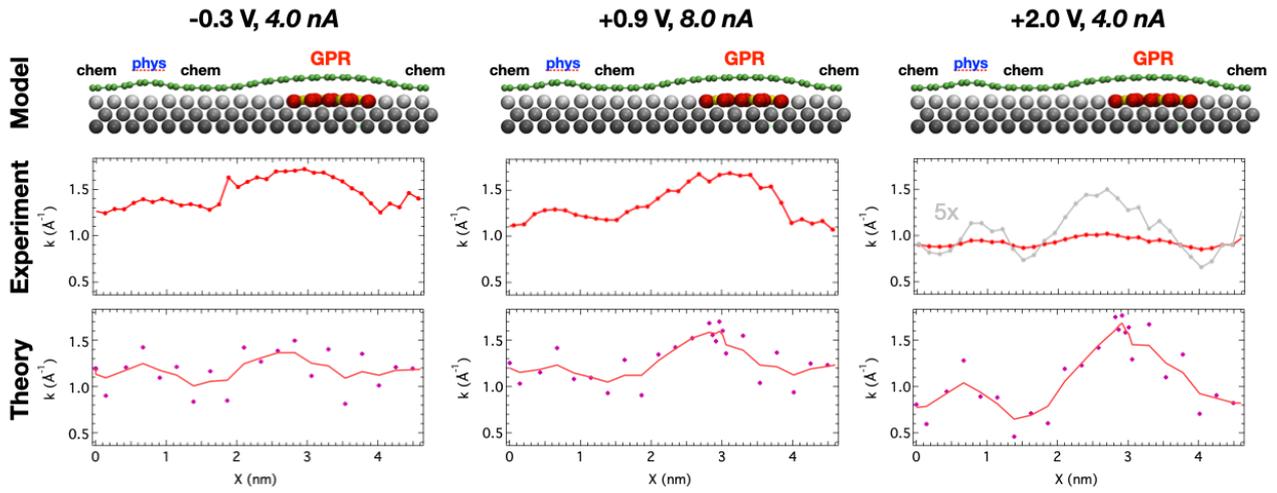

*Figure 3 (2-column, color) : Tunneling current decay k values at different lateral positions. From top to bottom: side view of the model cell to guide the eye; experimental decay constants of the tunneling current for selected points across the cell, collected at three different bias values (the tunneling currents at the starting point are written in italic); theoretical decay constants as calculated (points) and after a mild gaussian smoothing (in red), evaluated at the corresponding energy. The experimental plot at bias +2.0 V is also magnified 5X (in gray) to highlight its modulation.*

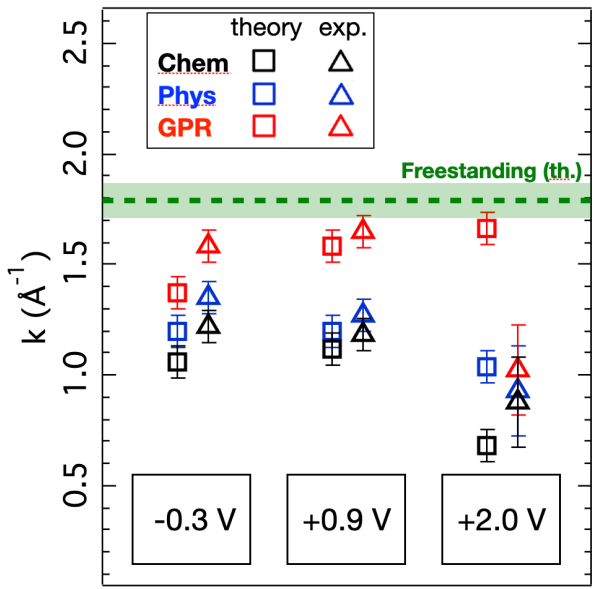

*Figure 4 (1-column, color): bias-dependency of tunneling current decay. Experimental and calculated I(z) decay constants for three different bias/energy values. Each experimental value is extracted by averaging the fit results of 50 decay spectra. The theoretical value of a flat,*

*freestanding graphene sheet is displayed as a green dashed line. Calculation error is displayed as a light green area.*

4.3 Field Emission Resonance

Since the tunneling current has an explicit dependence on the sample work function through the WKB transmission function, we performed Field Emission Resonance (FER) measurements to determine the effective work function shifts between the three considered graphene areas, and to gain further insight in the physical mechanism yielding the tunneling current decay constant *k*. The FER technique analyses the Gundlach oscillations in the *dz/dV* spectra, produced by the standing-wave electronic states in the tip-sample gap[4,25]. The work function shift between adjacent areas can be extracted, with a lateral resolution of ~ 1 nm[26], from the relative energy shift of the high-order oscillations[27]. Such method has already been successfully used on nanostructured graphene, e.g. for investigating the moiré-modulated interaction of graphene on Rh(111)[7] and on Ru(0001)[28], or the effect of Co intercalation on the electronic properties of graphene grown on Pt(111) surface[29]. **Figure 5** presents typical *dz/dV* spectra measured above chemisorbed (black)/physisorbed (blue) strongly interacting graphene and above GPR (red), coexisting on the Ni(100) surface. Low-energy resonances (2-4 V) are typically generated by the electron states localized between the graphene layer and the substrate[7,29]. Gundlach resonances at higher bias values are labeled with ordinal number *n.* For n = 1,2 the peak shifts correspond to the local work function shift[27]. Therefore, the work function shifts by 0.12 eV between chemisorbed and physisorbed lanes of the strong interacting graphene, while between physisorbed lanes and GPRs the shift is 0.36 eV. Therefore, the work function for graphene at the GPR center results from 0.36 eV to 0.48 eV higher than the work function of strongly interacting, moiré-modulated graphene on clean Ni(100). This is also in very good agreement with our DFT simulations, predicting a work function difference between chemisorbed graphene and noninteracting graphene of ~ 0.5 eV[30].

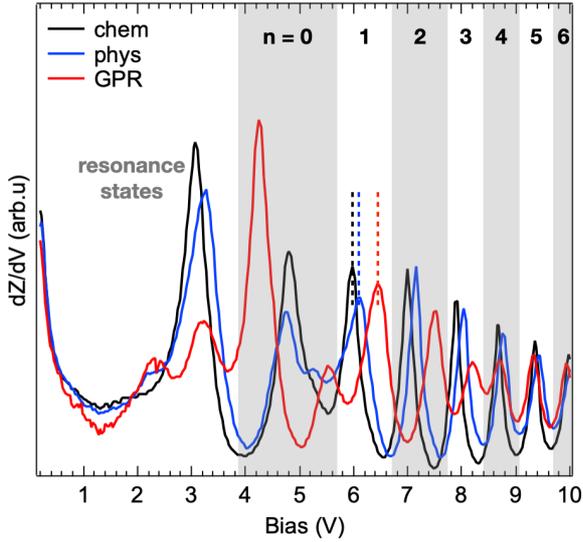

*Figure 5 (1-column, color): FER shifts. Field Emission Resonance plot measured above the regions indicated in the model cell shown in Figure 1. The low-energy resonance states and the proper Gundlach oscillations with ordinal number n are highlighted. The peak positions, relatively shifted by 0.12 and 0.36 eV in the n=1 resonance, are highlighted by three dashed lines. Spectra were acquired with lock-in oscillation amplitude 20 mV and frequency 1015 Hz.*

5. DISCUSSION

We showed above that *I(z)* spectroscopy detects substantial changes in the tunneling current exponential decay constant measured above strongly and weakly interacting graphene. Such variations of *k* can be ascribed to combined chemical and morphologic effects, the former due to the different interaction with the substrate, affecting the local DOS and the work function, and the latter due to purely geometric effects, i.e. the graphene curvature.

The two contributions can be separated in the DFT simulations. In particular, the morphologic effect can be investigated by modelling a freestanding sheet of graphene forced to keep the same undulated morphology of the graphene/Ni(100) model depicted in Figure 1b. The tunneling current decay constant *k* calculated in the regions with different curvature remains the same within the numerical accuracy, i.e. its modulation is one order of magnitude smaller than the one calculated and measured when graphene interacts with the substrate. Therefore, morphology cannot be the main cause of the witnessed *k* changes, that have to be necessarily ascribed to the different interaction with the substrate.

Experimentally, the dependance of *k* respect to the tip/sample distance can be evaluated by comparing the tunneling current decay extracted with different fit ranges. Ruling out the morphologic effect, the value *of k* may depend on the energy range and on the spatial extension of the electronic orbitals involved in the tunneling process. The influence on *k* of the latter for a given energy range can be estimated by shifting the 100 pm fit range towards the tail of the current decay. Figure S1 shows the *k* values across the cell (as in Figure 3) extracted by shifting the fit range by 0, 50, 100 and 150 pm, respectively. The shift causes a progressive underestimation of the fit values, but the uniform damping of the differences between distinct areas proves that there is no significant variation of the orbital extension along *z*. It must also be noticed that this effect does not imply that *I(z)* spectroscopy must be carried out as close to the surface as possible. In fact, Figure S1 shows that if the tip is too close, the extracted *k* values can be affected e.g. by a mechanic interaction between the pushing tip and the suspended GPR. Nonetheless, we emphasize the fact that the best conditions for evaluating the tunneling decay differences are produced with low bias (absolute value below ~ 1 V) and high tunneling current (several nA). Such conditions are quantitatively comparable with a theoretical distance from the topmost carbon atom in graphene of few Å. As seen for bias + 2.0 V, a tip placed too far away senses only the tail of the decay and gives a systematic underestimation of the effect.

The considerations expressed so far support the idea that *I(z)* can effectively sense the variations in the graphene DOS induced by the interaction with the substrate. In the following, we try to elucidate in more detail the physical relationship between the tunneling current decay and the graphene DOS. In general, the interaction strength between graphene and transition metal substrates can vary. The main discriminant factor is the abundance of electronic states close to Fermi energy, which can be quantified in first approximation by the energy position of the d-band centroid[15]: metals with d-band centroid binding energy smaller than ~ 2 eV (Pd, Rh, Ru, Ni, Co, Re) strongly interact with graphene, while the others (Pt, Au, Ir, Ag, Cu) interact only weakly. In the latter case, the band structure of graphene is mostly preserved: the linearly dispersed π-band is almost rigidly shifted in energy by few tenths of eV because of the charge redistribution at the graphene-metal interface[24]. In these weakly interacting systems, the graphene work function changes from the freestanding value of 4.5 eV to a range between 4.85 eV (Pt) and 4.4 eV (Cu)[31–33]. Conversely, on strongly interacting substrates the charge reordering at the graphene/metal interface lowers the graphene work function more significantly, ranging from 4.3 eV (Pd, Rh) to 4.0 eV (Ni)[7,32,34,35]. The different work function range between strongly (4.0-4.3 eV) and weakly (4.4-4.85 eV) interacting graphene is exactly the one sensed by FER in the system presented above: the measured 0.36-0.48 eV shift and the work function value measured for strongly interacting graphene on nickel (~ 4.0 eV) allows us to estimate a work function value of ~ 4.4 for graphene on interfacial nickel carbide, i.e. in the range of weakly interacting graphene. This work function difference, created by the different charge distribution, influences also the tunneling current through the transmission function reported in Eq. 2, thus changing the $I(z)$ decay constant. The reason why interfacial nickel carbide can locally deactivate the graphene/substrate interaction is explained in more detail through Figure S2. Panel (a) displays the calculated projected DOS of the nickel d-orbitals for a clean Ni(100) surface (black) and for $Ni_2C$(100) (red). The evident DOS depletion close to Fermi energy caused by the carbon atoms at the interface shifts the d-band centroid towards higher binding energies, from 1.07 eV to 1.51 eV, i.e. still below the empirical threshold of ~ 2 eV described before. This finding enlightens one intrinsic limit of such indicator for the graphene/substrate interaction. For interfaces more complex than a homonuclear, low-index metallic surface, the DOS depletion around Fermi energy seems to be a more appropriate descriptor. Moreover, panel (b) depicts the difference in the electron distribution between the overall graphene/Ni(100) system and its separate components (graphene and Ni(100) slab), frozen in their original positions. Most of the changes occur in the chemisorbed area, where C $p_z$ and Ni $d_{xz}$, $d_{yz}$ orbitals increase their charge, and Ni $d_{z^2}$ orbitals suffer a charge depletion. This binding mechanism, similar to the one occurring between graphene and Ni(111)[36,37], is deactivated by the interfacial carbon atoms that hybridize with the Ni $d$ orbitals to form nickel carbide, thus "making them unavailable" for further bonding with graphene.

Finally, the question arises if this investigation method can be reliably extended to other substrates. As benchmark, we applied our approach also to the buckled phase of graphene on Ir(100)[11]. This system presents a quasi-freestanding graphene, occasionally forced by the 1D moiré pattern to strongly interact with the substrate. Therefore, this configuration is to some respect the counterpart of graphene on Ni(100), where graphene is instead generally strongly interacting and just occasionally lifted to form GPRs. Figure S3 shows STM topography, the atomic model used for DFT calculation, experimental I(z) spectra and FER resonances of the buckled phase of graphene on Ir(100). We collected tunneling decay spectra with the same bias and initial starting tunneling current used for the Ni(100) case, and extracted the decay constant according to the same procedure. For bias +0.9 V we obtained 1.50 Å$^{-1}$ (non-interacting) and 1.28 Å$^{-1}$ (interacting), while, for bias -0.3 V, 1.44 Å$^{-1}$ (non-interacting) and 1.07 Å$^{-1}$ (interacting). Also, for this case we performed DFT calculations, obtaining 1.65 Å$^{-1}$ (non-interacting) and 1.24 Å$^{-1}$ (interacting) for bias +0.9 V, while 1.47 Å$^{-1}$ (non-interacting) and 1.41 Å$^{-1}$ (interacting) for bias -0.3 V. Experimental and numerical uncertainty of ±0.07 Å$^{-1}$ can be attributed to all values. The agreement between DFT predictions and the measured values is in general excellent, if we consider also the difficulty of uniquely identifying the small regions of graphene interacting with Ir(100), that could be a source of the major discrepancy between theory and experiment for bias -0.3 V. Overall, the *k* values for graphene on Ir(100) are perfectly comparable with the ones presented for the topical case of nanostructured graphene on Ni(100). Therefore, we are highly confident that the proposed method can be in principle extended to many other cases, including virtually any other substrate. On the basis of our proof of concept and specifically on the nanometer-size probed areas, one may even consider extending our method to systems where nanostructured modification of graphene/substrate interaction is driven by single metal atoms, small molecules or clusters trapped underneath graphene, although more specific investigations would be needed.

## 6. CONCLUSION

In summary, *I(z)* spectroscopy is shown to be a simple method to distinguish the kind of substrate interaction occurring in a nanostructured graphene grown on Ni(100). The tunneling current decay undergoes substantial changes when measured over a strongly or weakly interacting graphene region: its exponential decay constant decreases as the graphene/substrate interaction strengthens, and can range from a value close to the one calculated for freestanding graphene (1.79±0.07 Å$^{-1}$) to about 1.1-1.2 Å$^{-1}$ for chemisorbed graphene. Purely morphological effects due to the graphene corrugation can be ruled out on the basis of DFT simulations. This variability can be connected to the changes in the graphene work function, experimentally evidenced by FER spectroscopy and induced by the strength of the bonding between the graphene π orbitals and the substrate *d* orbitals. Therefore, *I(z)* spectroscopy can be used to quantitatively describe the degree of interaction between graphene and a metal substrate. DFT calculations of the tunneling decay constant and of the work function in different conditions of graphene/substrate interaction fully support the reliability of the information extracted from the experiments. Furthermore, this method was tested on graphene/Ir(100), where the buckled phase locally forces a stronger graphene/substrate interaction. Again, I(z) experimental and theoretical findings agree quantitatively. More generally, this result demonstrates that reliable spectroscopic information on graphene can be retrieved by simple STM measurements and that *I(z)* can be fairly used on a generic graphene/substrate system to extend the STM experimental findings beyond topographic imaging.

ACKNOWLEDGEMENTS

The authors thank Prof. Silvio Modesti for fruitful discussion. A.S. and M. Pe. acknowledge support from the Italian Ministry of University and Research (MUR) through the program PRIN 2017 – Project no. 2017KFY7XF. M.Pa. and C.A. acknowledge support from the Italian Ministry of University and Research (MUR) through the program PRIN 2017 – Project no. 2017NYPHN8. We acknowledge access to the Cineca high performance computing resources through the agreement with the University of Trieste and ISCRA projects.

REFERENCES

[1] R.M. Feenstra, J.A. Stroscio, A.P. Fein, Tunneling spectroscopy of the Si(111)2 × 1 surface, Surf Sci. 181 (1987) 295–306. https://doi.org/10.1016/0039-6028(87)90170-1.

[2] R.M. Feenstra, Scanning tunneling spectroscopy, Surf Sci. 299–300 (1994) 965–979. https://doi.org/10.1016/0039-6028(94)90710-2.

[3] R.S. Becker, J.A. Golovchenko, B.S. Swartzentruber, Electron Interferometry at Crystal Surfaces, Phys Rev Lett. 55 (1985) 987–990. https://doi.org/10.1103/PhysRevLett.55.987.

[4] K.H. Gundlach, Zur berechnung des tunnelstroms durch eine trapezförmige potentialstufe, Solid State Electron. 9 (1966) 949–957. https://doi.org/10.1016/0038-1101(66)90071-2.

[5] L. Zhao, R. He, K.T. Rim, T. Schiros, K.S. Kim, H. Zhou, C. Gutiérrez, S.P. Chockalingam, C.J. Arguello, L. Pálová, D. Nordlund, M.S. Hybertsen, D.R. Reichman, T.F. Heinz, P. Kim, A. Pinczuk, G.W. Flynn, A.N. Pasupathy, Visualizing individual nitrogen dopants in monolayer graphene., Science (1979). 333 (2011) 999–1003. https://doi.org/10.1126/science.1208759.

[6] S. Fiori, D. Perilli, M. Panighel, C. Cepek, A. Ugolotti, A. Sala, H. Liu, G. Comelli, C. Di Valentin, C. Africh, "Inside out" growth method for high-quality nitrogen-doped graphene, Carbon N Y. 171 (2021) 704–710. https://doi.org/10.1016/j.carbon.2020.09.056.

[7] B. Wang, M. Caffio, C. Bromley, H. Früchtl, R. Schaub, Coupling Epitaxy, Chemical Bonding, and Work Function at the Local Scale in Transition Metal-Supported Graphene, ACS Nano. 4 (2010) 5773–5782. https://doi.org/10.1021/nn101520k.

[8] A. Sala, Z. Zou, V. Carnevali, M. Panighel, F. Genuzio, T.O. Menteş, A. Locatelli, C. Cepek, M. Peressi, G. Comelli, C. Africh, Quantum Confinement in Aligned Zigzag "Pseudo-Ribbons" Embedded in Graphene on Ni(100), Adv Funct Mater. 32 (2022) 2105844. https://doi.org/10.1002/adfm.202105844.

[9] Z. Zou, V. Carnevali, M. Jugovac, L.L. Patera, A. Sala, M. Panighel, C. Cepek, G. Soldano, M.M. Mariscal, M. Peressi, G. Comelli, C. Africh, Graphene on nickel (100) micrograins: Modulating the interface interaction by extended moiré superstructures, Carbon N Y. 130 (2018) 441–447. https://doi.org/10.1016/J.CARBON.2018.01.010.

[10] Z. Zou, L.L. Patera, G. Comelli, C. Africh, Strain release at the graphene-Ni(100) interface investigated by in-situ and operando scanning tunnelling microscopy, Carbon N Y. 172 (2021) 296–301. https://doi.org/10.1016/J.CARBON.2020.10.019.

[11] A. Locatelli, C. Wang, C. Africh, N. Stojić, T.O. Menteş, G. Comelli, N. Binggeli, Temperature-driven reversible rippling and bonding of a graphene superlattice., ACS Nano. 7 (2013) 6955–63. https://doi.org/10.1021/nn402178u.

[12] P. Giannozzi, S. Baroni, N. Bonini, M. Calandra, R. Car, C. Cavazzoni, D. Ceresoli, G.L. Chiarotti, M. Cococcioni, I. Dabo, A. Dal Corso, S. de Gironcoli, S. Fabris, G. Fratesi, R. Gebauer, U. Gerstmann, C. Gougoussis, A. Kokalj, M. Lazzeri, L. Martin-Samos, N. Marzari, F. Mauri, R. Mazzarello, S. Paolini, A. Pasquarello, L. Paulatto, C. Sbraccia, S. Scandolo, G. Sclauzero, A.P. Seitsonen, A. Smogunov, P. Umari, R.M. Wentzcovitch, QUANTUM ESPRESSO: a modular and open-source software project for quantum simulations of materials, Journal of Physics: Condensed Matter. 21 (2009) 395502. https://doi.org/10.1088/0953-8984/21/39/395502.

[13] J.P. Perdew, K. Burke, M. Ernzerhof, Generalized Gradient Approximation Made Simple, Phys Rev Lett. 77 (1996) 3865–3868. https://doi.org/10.1103/PhysRevLett.77.3865.

[14] S. Grimme, Density functional theory with London dispersion corrections, WIREs Computational Molecular Science. 1 (2011) 211–228. https://doi.org/10.1002/wcms.30.

[15] A. Dahal, M. Batzill, Graphene–nickel interfaces: a review, Nanoscale. 6 (2014) 2548. https://doi.org/10.1039/c3nr05279f.

[16] J.G. Simmons, Generalized Formula for the Electric Tunnel Effect between Similar Electrodes Separated by a Thin Insulating Film, J Appl Phys. 34 (1963) 1793–1803. https://doi.org/10.1063/1.1702682.

[17] V.A. Ukraintsev, Data evaluation technique for electron-tunneling spectroscopy, Phys Rev B. 53 (1996) 11176–11185. https://doi.org/10.1103/PhysRevB.53.11176.

[18] M. Ziegler, N. Néel, A. Sperl, J. Kröger, R. Berndt, Local density of states from constant-current tunneling spectra, Phys Rev B. 80 (2009) 125402. https://doi.org/10.1103/PhysRevB.80.125402.

[19] A. Pronschinske, D.J. Mardit, D.B. Dougherty, Modeling the constant-current distance-voltage mode of scanning tunneling spectroscopy, Phys Rev B. 84 (2011) 205427. https://doi.org/10.1103/PhysRevB.84.205427.

[20] M. Passoni, F. Donati, A. Li Bassi, C.S. Casari, C.E. Bottani, Recovery of local density of states using scanning tunneling spectroscopy, Phys Rev B. 79 (2009) 045404. https://doi.org/10.1103/PhysRevB.79.045404.

[21] B. Koslowski, C. Dietrich, A. Tschetschetkin, P. Ziemann, Evaluation of scanning tunneling spectroscopy data: Approaching a quantitative determination of the electronic density of states, Phys Rev B. 75 (2007) 035421. https://doi.org/10.1103/PhysRevB.75.035421.

[22] C. Hellenthal, R. Heimbuch, K. Sotthewes, E.S. Kooij, H.J.W. Zandvliet, Determining the local density of states in the constant current STM mode, Phys Rev B. 88 (2013) 035425. https://doi.org/10.1103/PhysRevB.88.035425.

[23] J. Tersoff, D.R. Hamann, Theory of the scanning tunneling microscope, Phys Rev B. 31 (1985) 805–813. https://doi.org/10.1103/PhysRevB.31.805.

[24] G. Giovannetti, P.A. Khomyakov, G. Brocks, V.M. Karpan, J. van den Brink, P.J. Kelly, Doping Graphene with Metal Contacts, Phys Rev Lett. 101 (2008) 026803. https://doi.org/10.1103/PhysRevLett.101.026803.

[25] G. Binnig, K.H. Frank, H. Fuchs, N. Garcia, B. Reihl, H. Rohrer, F. Salvan, A.R. Williams, Tunneling Spectroscopy and Inverse Photoemission: Image and Field States, in: Springer, Dordrecht, 1985: pp. 93–96. https://doi.org/10.1007/978-94-011-1812-5_11.

[26] P. Ruffieux, K. Aït-Mansour, A. Bendounan, R. Fasel, L. Patthey, P. Gröning, O. Gröning, Mapping the Electronic Surface Potential of Nanostructured Surfaces, Phys Rev Lett. 102 (2009) 086807. https://doi.org/10.1103/PhysRevLett.102.086807.

[27] C.L. Lin, S.M. Lu, W.B. Su, H.T. Shih, B.F. Wu, Y.D. Yao, C.S. Chang, T.T. Tsong, Manifestation of Work Function Difference in High Order Gundlach Oscillation, Phys Rev Lett. 99 (2007) 216103. https://doi.org/10.1103/PhysRevLett.99.216103.

[28] D. Niesner, T. Fauster, Image-potential states and work function of graphene, Journal of Physics: Condensed Matter. 26 (2014) 393001. https://doi.org/10.1088/0953-8984/26/39/393001.

[29] M. Gyamfi, T. Eelbo, M. Waśniowska, R. Wiesendanger, Impact of intercalated cobalt on the electronic properties of graphene on Pt(111), Phys Rev B. 85 (2012) 205434. https://doi.org/10.1103/PhysRevB.85.205434.

[30] We calculated the work function of epitaxial graphene on Ni(111) at its equilibrium distance (chemisorbed configuration) and at a larger distance (noninteracting configuration), obtaining 4.17 eV and 4.66 eV, respectively, i.e. a difference of 0.49 eV. The two configurations, although morphologically different from the chemisorbed stripes and the GPR in nanostructured graphene on Ni(100), are well representative of the strong and weak interaction regime, respectively, and, since they are described by the same simulation cell, the numerical uncertainty of their work function difference is minimal.

[31] P.A. Khomyakov, G. Giovannetti, P.C. Rusu, G. Brocks, J. van den Brink, P.J. Kelly, First-principles study of the interaction and charge transfer between graphene and metals, Phys Rev B. 79 (2009) 195425. https://doi.org/10.1103/PhysRevB.79.195425.

[32] D. Nobis, M. Potenz, D. Niesner, T. Fauster, Image-potential states of graphene on noble-metal surfaces, Phys Rev B. 88 (2013) 195435. https://doi.org/10.1103/PhysRevB.88.195435.

[33] B. Cook, A. Russakoff, K. Varga, Coverage dependent work function of graphene on a Cu(111) substrate with intercalated alkali metals, Appl Phys Lett. 106 (2015) 211601. https://doi.org/10.1063/1.4921756.

[34] C. Oshima, A. Nagashima, Ultra-thin epitaxial films of graphite and hexagonal boron nitride on solid surfaces, Journal of Physics: Condensed Matter. 9 (1997) 1–20. https://doi.org/10.1088/0953-8984/9/1/004.

[35] N. Armbrust, J. Güdde, P. Jakob, U. Höfer, Time-Resolved Two-Photon Photoemission of Unoccupied Electronic States of Periodically Rippled Graphene on Ru(0001), Phys Rev Lett. 108 (2012) 056801. https://doi.org/10.1103/PhysRevLett.108.056801.

[36] A. Varykhalov, D. Marchenko, J. Sánchez-Barriga, M.R. Scholz, B. Verberck, B. Trauzettel, T.O. Wehling, C. Carbone, O. Rader, Intact Dirac Cones at Broken Sublattice Symmetry: Photoemission Study of Graphene on Ni and Co, Phys Rev X. 2 (2012) 041017. https://doi.org/10.1103/PhysRevX.2.041017.

[37] F. Bianchini, L.L. Patera, M. Peressi, C. Africh, G. Comelli, Atomic Scale Identification of Coexisting Graphene Structures on Ni(111), J Phys Chem Lett. 5 (2014) 467–473. https://doi.org/10.1021/jz402609d.

AUTHOR CONTRIBUTION

Virginia Carnevali: Investigation, Methodology, Software, Validation, Visualization, Writing – review and editing; Alessandro Sala: Investigation, Methodology, Formal analysis, Data curation, Writing – original draft; Pietro Biasin: Investigation; Mirco Panighel: Investigation, Methodology; Giovanni Comelli: Funding acquisition, Resources, Writing – review and editing; Maria Peressi: Conceptualization, Funding acquisition, Resources, Formal analysis, Writing – review and editing; Cristina Africh: Conceptualization, Funding acquisition, Resources, Writing – review and editing.

# SUPPORTING INFORMATION

## Probing the graphene/substrate interaction by electron tunneling decay

*V. Carnevali [a,1], A. Sala [a,b,*], P. Biasin [a], M. Panighel [b], G. Comelli [a,b], M. Peressi [a,*], C. Africh [b]*

[a] *Department of Physics, University of Trieste, via Valerio 2, 34127, Trieste, Italy*
[b] *CNR-IOM, Laboratorio TASC, S.S. 14 km 163.5, Basovizza, 34149, Trieste, Italy*

\* Corresponding authors: A. Sala sala@iom.cnr.it (experiment), M. Peressi peressi@units.it (theory)
[1] Present address: Institute of Chemical Science and Engineering, EPFL, 1015 Lausanne, Switzerland

**Table of content**

**Figure S1**

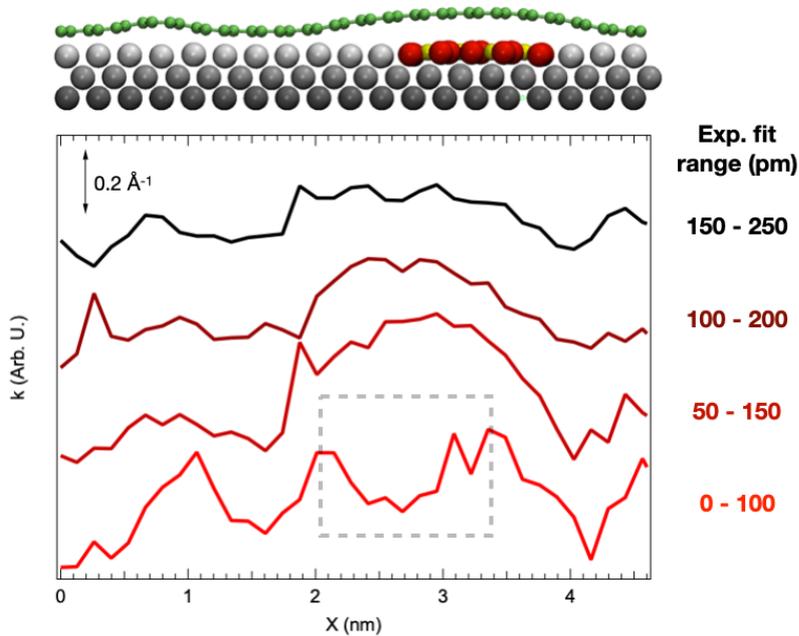

*Figure S1 displays the cascade plot of the experimental I(z) decay constant for four different fit ranges (spelled out on the right). The closest fit range, 0-100 pm, underestimates systematically the decay constant at the GPR center (inside the dashed grey box) because of a strong mechanical interaction between probe tip and lifted graphene. Farther fit ranges return progressively reduced values of the decay constant. Profiles extracted with bias -0.3 V. The profile with range 50-150 is the one displayed in Figure 3.*

**Figure S2**

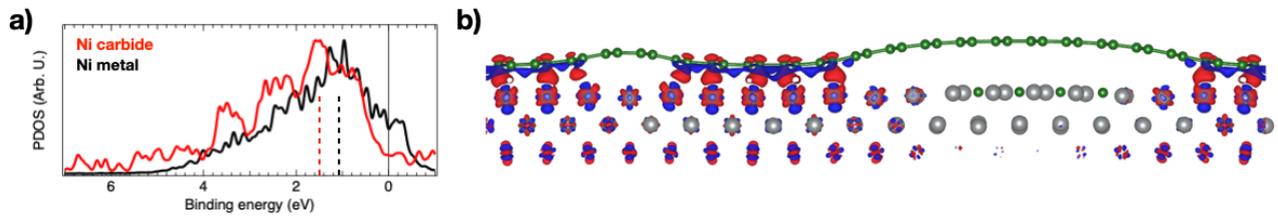

*a) Projected DOS of Ni surface atoms in metallic state (black) and in $Ni_2C$ (red). The corresponding values of d-band centers are indicated by vertical colored dashed lines.*
*b) Electron density difference between the overall graphene/Ni(100) system and its separate components (graphene and Ni(100) slab), frozen in their original positions: differential density isosurfaces, set at ±0.004 |e|/a.u.³ for positive (red) and negative (blue) values, displayed from the side.*

**Figure S3**

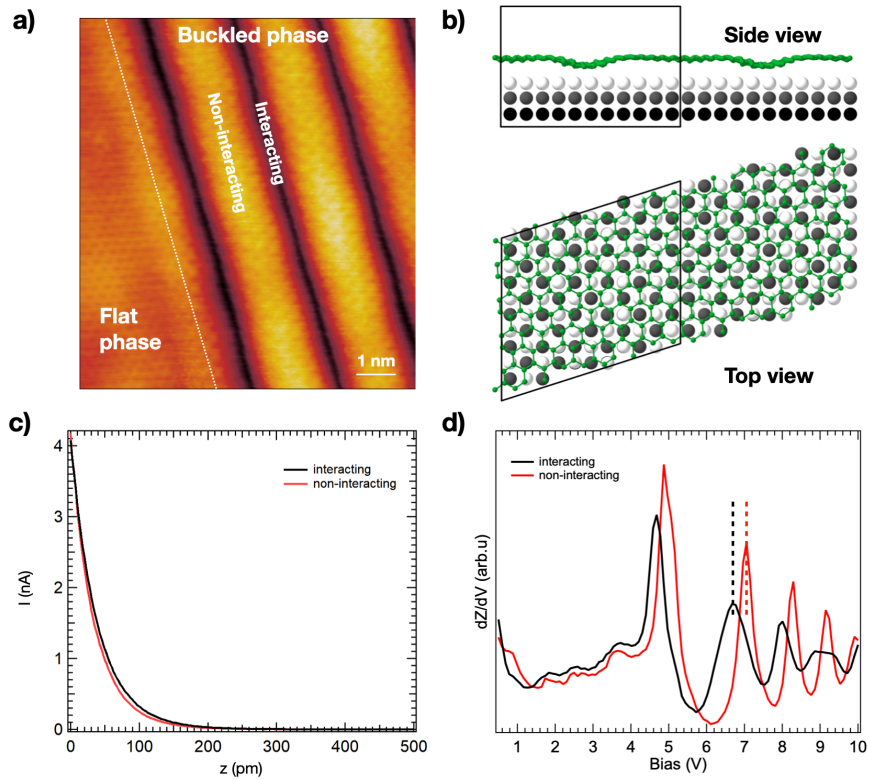

*a) STM topographic image of graphene grown via CVD of ethylene on Ir(100). Bias -0.3 V, tunnelling current 4.0 nA. Graphene presents two separate phases, one flat on the surface with quasi-freestanding conditions, and one buckled with 1D wavy corrugation of lateral periodicity ˜2.1 nm. In the latter, the ridges constitute of quasi-freestanding graphene, while in the valleys the graphene is strongly interacting with the substrate. More information on this system can be found in Ref. [11]. b) ball rendering of the supercell used for DFT calculations, containing 3 Ir layers with 55 atoms each one and 160 C atoms, after the optimization of the atomic positions. Graphene is depicted in green, while the Ir atoms are displayed in grey shades. The graphene distance from the Ir surface ranges from 2.3 Å to 3.2 Å. For the calculations of the density of states and the decay constant, a supercell with a larger vacuum spacing (20 Å) has been used. c) Plot of the tunneling current decay along the z direction (surface normal) obtained by positioning the tip on interacting and non-interacting lanes of the buckled phase at bias +0.9 V. The difference in the decay rate is visible. The decay constant values are 1.44 Å$^{-1}$ (non-interacting) and 1.07 Å$^{-1}$ (interacting). d) Field Emission Resonance plot measured above the interacting and non-interacting lanes of the buckled phase. The first-order peak positions, relatively shifted by 0.35 eV, are highlighted by dashed lines. Spectra were acquired with lock-in oscillation amplitude 20 mV and frequency 970 Hz.*